\documentclass[sigconf, 10pt]{acmart}

\AtBeginDocument{%
  }
\setcopyright{acmlicensed}
\copyrightyear{2018}
\acmYear{2018}
\acmDOI{XXXXXXX.XXXXXXX}

\acmConference[MobiArch '24]{19th  Workshop on Mobility in the Evolving Internet Architecture}{18th Nov, 2024}{Washington, D.C., USA}
\acmBooktitle{MobiArch '24: Proceedings of the 19th Workshop on Mobility in the Evolving Internet Architecture,
18 November 2024, Washington, DC, USA}
\renewcommand\footnotetextcopyrightpermission[1]{}
\usepackage{float}%提供float浮动环境
\usepackage{booktabs}%提供命令\toprule、\midrule、\bottomrule
\usepackage{bbding} 
\usepackage{xspace}
\usepackage{verbatim}
\usepackage{amsfonts}
\usepackage{amsmath}

\usepackage{enumerate}
\usepackage{threeparttable}
\usepackage{tabularx}

\begin{document}
\title{A First Look At Efficient And Secure On-Device LLM Inference Against KV Leakage}

\author{Huan Yang}
% \authornote{None}
\email{yanghuan9812@csu.edu.cn}
\affiliation{%
  \institution{Central South University}
  \city{Changsha}
  \state{Hunan}
  \country{China}
}

\author{Deyu Zhang}
\authornote{Deyu Zhang (zdy876@csu.edu.cn) is the corresponding author.}
\email{zdy876@csu.edu.cn}
\affiliation{%
  \institution{Central South University}
  \city{Changsha}
  \state{Hunan}
  \country{China}
}

\author{Yudong Zhao}
% \authornote{None}
\email{yudong.zhao@transsion.com}
\affiliation{%
  \institution{Shanghai Transsion CO., LTD}
    \city{Shanghai}
  % \state{Shanghai}
  \country{China}
}

\author{Yuanchun Li}
% \authornote{None}
\email{liyuanchun@air.tsinghua.edu.cn}
\affiliation{%
  \institution{Tsinghua University}
\city{Beijing}
  % \state{Beijing}
  \country{China}
}

\author{Yunxin Liu}
\email{liuyunxin@air.tsinghua.edu.cn}
\affiliation{%
  \institution{Tsinghua University}
\city{Beijing}
  % \state{Beijing}
  \country{China}
}

\begin{abstract}
Running LLMs on end devices has garnered significant attention recently due to their advantages in privacy preservation. With the advent of lightweight LLM models and specially designed GPUs, on-device LLM inference has achieved the necessary accuracy and performance metrics.

However, we have identified that LLM inference on GPUs can leak privacy-sensitive intermediate information, specifically the KV pairs. An attacker could exploit these KV pairs to reconstruct the entire user conversation, leading to significant vulnerabilities. Existing solutions, such as Fully Homomorphic Encryption (FHE) and Trusted Execution Environments (TEE), are either too computation-intensive or resource-limited.

To address these issues, we designed KV-Shield, which operates in two phases. In the initialization phase, it permutes the weight matrices so that all KV pairs are correspondingly permuted. During the runtime phase, the attention vector is inversely permuted to ensure the correctness of the layer output. All permutation-related operations are executed within the TEE, ensuring that insecure GPUs cannot access the original KV pairs, thus preventing conversation reconstruction. Finally, we theoretically analyze the correctness of KV-Shield, along with its advantages and overhead.
% Running LLM on end devices has received extensive attentions recently due to its advantages in privacy preserving. Given the emerging of the lightweight LLM model and specially designed AI accelerators, the on-device LLM inference has passed the accuracy and performance metrics.

% However, we find that the LLM inference on AI-accelerators can leak the privacy-sensitive intermediate information, i.e., the KV pairs. Based on the KV pairs, an attacker can reconstruct the entire user conversation, leading to significant vulnerabilities. We further demonstrate that the existing solutions, such as Fully Homomorphic Encryption (FHE) and Trusted Execution Environment (TEE), are either too computation-intensive or resource-limited. Addressing the issues, we design KV-Shield that runs in two phases. In the initialization phase, it permutes the weight matrices, such that all the KV pairs will be correspondingly permuted. In the runtime phase, the attention vector is reversely permuted to guarantee the correctness of the layer output. All the permutation-related operations are accomplished in the TEE, such that the insecure GPUs cannot access the original KV pairs, thus preventing conversation recreation. Finally, we theoretically analyze the correctness of KV-Shield, and its advantages and overhead.
\end{abstract}
\maketitle

\section{Introduction}
% 自从2023年OpenAI推出ChatGPT的大语言模型服务以来，各种基于Transfomer结构的LLM模型迭代发展迅速，一年内Meta将LLaMA\cite{touvron2023llama} 迭代到了LLaMA3。得益于语言模型的强大，各类AI助手，例如代码助手，智能手机助手类的应用可以帮助生产者们大幅度提升工作效率。与此同时，开发者们正在想方设法地将LLM部署到各类移动设备上，例如ARM CPU、ARM GPU等，甚至一些NPU制造商们也在尝试使用NPU运行Transformer模型。
% 对于桌面端计算芯片和移动端计算芯片，例如Nvidia GPU、ARM GPU、Snapdragon GPU，可能由于一些设计问题它们通常都会存在一些潜在的风险漏洞。一些工作\cite{mittal2021survey},\cite{mittal2018survey}对于现有的GPU等硬件进行了漏洞风险的评估,硬件上存在的风险可能会导致商用且闭源的模型架构和权重被窃取,甚至用户的推理结果被恶意干扰和修改。
% Leftoverlocal\cite{sorensen2024leftoverlocals}的作者发现，目前在一些移动设备上，存在GPU内存漏洞，会使得共享内存的数据泄漏。通过他们的实验，成功的在AMD的GPU设备上通过漏洞窃取了用户和LLM的对话内容。我们同样在Xiaomi 13移动手机上成功的复现了该漏洞，成功的获取到了LLM和用户的部分对话内容。（补充对话图片）
% 同时，目前移动端芯片，例如ARM CPU芯片，它们配备了Trusted Execution Environment(TEE)，用于保护用户的指纹、密码等重要数据。一些工作\cite{mo2020darknetz,sun2023shadownet}也已经研究过如何使用TEE保护CNN模型的推理安全性。相比与CNNs，LLMs的参数量数以十亿计，模型权重大小通常以GB为单位，例如ChatGLM3模型使用INT4精度量化，它仍需要6GB左右的内存占用。本文主要将深入的阐述基于Transformer结构的LLMs在端侧保护用户对话内容的挑战和问题。
Since the launch of the large language model (LLM) service by OpenAI in 2023, various LLM models based on the Transformer architecture have rapidly emerged. These models have enabled groundbreaking applications, such as expert-level programming and advanced smartphone assistants, poised to transform the information service access paradigm, much like search engines and operating systems did in the past.

Compared to transmitting privacy-sensitive data over the Internet, on-device execution of LLMs is considered the most privacy-preserving solution \cite{yin2024llmaas}. Current mobile device manufacturers are competitively releasing on-device LLM deployment solutions at both the software and hardware levels. Notable examples include Apple's nearly 3-billion-parameter model and Qualcomm's Snapdragon 8 Gen 3 NPU.

While on-device LLM inference has been validated for accuracy and efficiency, it now faces the critical test of security. Unfortunately, the computing cores of mobile devices are vulnerable to various attacks, particularly information leakage \cite{mittal2018survey}. For instance, the running kernel can be extracted from nearly all components of mobile GPUs, including shared, local, and texture memory. The impact of information leakage is magnified for LLMs. Leading LLM inference frameworks, like Meta's LLama \cite{touvron2023llama}, utilize memory caching of key-value (KV) pairs to accelerate inference. The KV cache persists throughout the entire inference process, lasting from seconds to minutes. Breaches in the KV cache can lead to the recreation of the original user conversation. A prime illustration is demonstrated in Leftoverlocal \cite{sorensen2024leftoverlocals} on an AMD GPU, where data leaks in shared memory allowed an attacker to intercept the KV cache and replicate the entire conversation. We have replicated this attack on a Xiaomi 12 equipped with a Snapdragon 8 Gen 1 SoC.

In this paper, we make the first effort to protect the KV cache saved in mobile memory during LLM inference from two perspectives:

\begin{itemize}
    \item \textbf{Making the KV cache uninformative}. To achieve this, we modify the KV pairs during LLM inference so that the original conversation cannot be recreated even if the KV pairs are leaked. We evaluate the performance of two solutions: Fully Homomorphic Encryption (FHE) and permutation.
    \item \textbf{Making the KV cache invisible}. We process the KV pairs in a Trusted Execution Environment (TEE), a secure area of the main processor commonly used in mainstream ARM architectures. This ensures that the KV cache is invisible to the outside insecure world.
\end{itemize}

We demonstrate that FHE is too computation-intensive for on-device LLM inference. The size of KV pairs is too large for the memory-limited TEE, which does not support GPU acceleration, significantly limiting the runtime performance of on-device LLM inference. Based on these insights, we design KV-Shield, which employs a lightweight encryption scheme, namely permutation, ensuring that insecure GPUs can only access the permuted KV pairs at runtime. Even if the permuted KV pairs are leaked, the user conversation cannot be reconstructed. We theoretically analyze the correctness of KV-Shield and discuss its overhead.

\section{Background}
% 本节将详细介绍现有移动端设备内存泄漏的风险并介绍现有对于大语言模型的内容保护方案。

% 现有的LLM推理系统，例如llama.cpp，通常着重于模型推理性能的优化，对于硬件和推理形式的漏洞通常只提供了安全性建议，它们无法有效地全方位的保护用户使用LLM过程中会话上下文的安全性。本章主要针对类似于llama.cpp这类开源的LLM推理引擎无法有效地保护用户对话内容(Key Value Cache)的现象进行分析和讨论。

This section details the risks of memory leaks on existing mobile devices and describes existing content protection schemes for large language models.

\subsection{Key Value Cache for LLMs} \label{sec:kv_for_llms}

% \deyu{Introduce how does the KV generated and why does the inference of LLM need to cache KV? Probably efficiency issue.}
% \huan{First, i introduce self-attention in transformer, explain how  attention generate kv. I will explain what Q, K, and V stand for and how they relate to the inputs and outputs of the attention module.}

The self-attention mechanism\cite{vaswani2017attention} in Transformer models is a core component used to capture dependencies between different positions in the input sequence. It flexibly focuses on different parts of the sequence to better understand the context.
% \deyu{no need to elobrate on the definition of QKV, just how they are generated.} \huan{I will introduce computation workflow of Self-Attention to show how QKV be generated}
\begin{equation}
\mathrm{Attention}(Q,K,V) = \mathrm{Softmax}(\frac{QK^{\top}}{\sqrt{d_{k}}})V
\label{eq:self_attention}
\end{equation}
As shown in Eq. \ref{eq:self_attention}, the input sequence is mapped through three weight matrices (linear transformations) to generate query (Q), key (K), and value (V) vectors. The attention scores are computed from the dot product of Q and K (where \(\sqrt{d_k}\) is the vector dimension), normalized using the softmax function, and then used to obtain a weighted sum of the V vectors, capturing long-range dependencies within the sequence.

Most large language models, such as LLaMA \cite{touvron2023llama} and Qwen \cite{bai2023qwen}, are built on the Causal Decoder architecture \cite{zhao2023survey}. These models generate tokens autoregressively, determining the next token based on the past prompt and previously generated tokens. Each time a new token's attention representation is computed, the corresponding Q-vector must be calculated with the K- and V-vectors of the past prompt and generated tokens. However, the K- and V-vectors for past prompts and generated tokens are already computed during previous generations.

Caching the K- and V-vectors in memory prevents redundant computation. Assuming the input sequence length is \( n \) and the model dimension is \( d_{\text{model}} \), the dimensions of the \(\mathbf{K}\) and \(\mathbf{V}\) matrices are \( n \times d_{\text{model}} \). Using cached \(\mathbf{K}\) and \(\mathbf{V}\) reduces the need for two matrix multiplications of size \([n-1, d] \times [d, d]\).

\begin{figure*}
\centering
\includegraphics[width=0.7\textwidth]{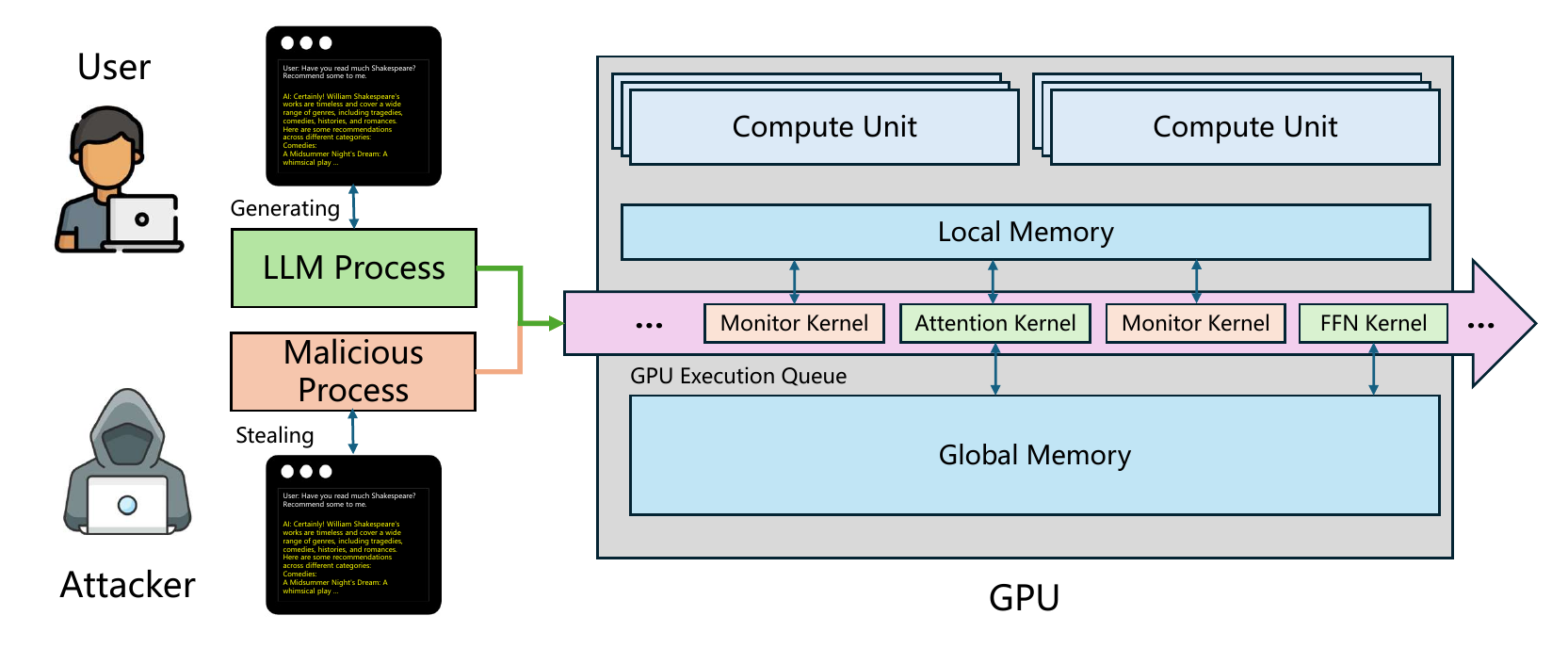}
\caption{Attacker's workflow for stealing user contexts. 
 }
\label{fig:attacker_workflow}
\end{figure*}

\subsection{Threat Model} \label{sec:recreate_from_user_kvcache}

\paragraph{\textbf{Scene Setting}}

We consider an adversary capable of observing GPU tasks in the normal world and exploiting vulnerabilities similar to Leftoverlocal\cite{sorensen2024leftoverlocals} to read the high-speed shared cache contents of the device GPU, such as OpenCL's local memory or CUDA's shared memory, but unable to directly access the GPU memory to obtain the model's inputs and outputs. Our primary goal is to protect the privacy of conversations between users and the large language model (LLM), rather than focusing on protecting the model weights. Such that the adversary cannot retrieve KV cache contents from the GPU's high-speed shared cache to reconstruct user conversations. Additionally, we do not consider side-channel attacks on the Trusted Execution Environment (TEE) — we assume the TEE can safeguard the confidentiality and integrity of its internal programs and data.

\paragraph{\textbf{How does the adversary reconstruct the user conversation?}}
As shown in Figure \ref{fig:attacker_workflow}, when the user initiates the LLM process, the attacker launches a malicious process to steal the user's KV cache. The LLM process continuously submits GPU tasks to the GPU execution queue, such as Attention Kernel and FFN Kernel, to perform model inference and generate tokens. The attacker's malicious process continuously generates monitor kernels and inserts them after the Attention Kernel. By exploiting vulnerabilities, it accesses the cached contents in the local memory of the Attention Kernel - the KV cache, and transmits this information to the attacker. The attacker can determine which open-source LLM is being used by analyzing the types of GPU tasks and the KV contents. Then, the attacker inputs the KV cache into the attention module along with a prompt provided by the attacker, thereby reconstructing the user's conversation content.

\section{Potential Solution Analysis} \label{sec:Potential_Solution_Analysis}

We analyzed two potential solutions, namely Fully Homomorphic Encryption (FHE) and running model inference in a Trusted Execution Environment (TEE), to prevent the recreation of user conversations caused by KV leakage. FHE encrypts the entire LLM inference process without affecting the accuracy of the model. TEE isolates the model inference process from the GPU and insecure memory. Running model inference in TEE makes intermediate results, such as KV pairs, invisible.

\subsection{The Performance of FHE}

\begin{figure}
\centering
\includegraphics[width=0.5\textwidth]{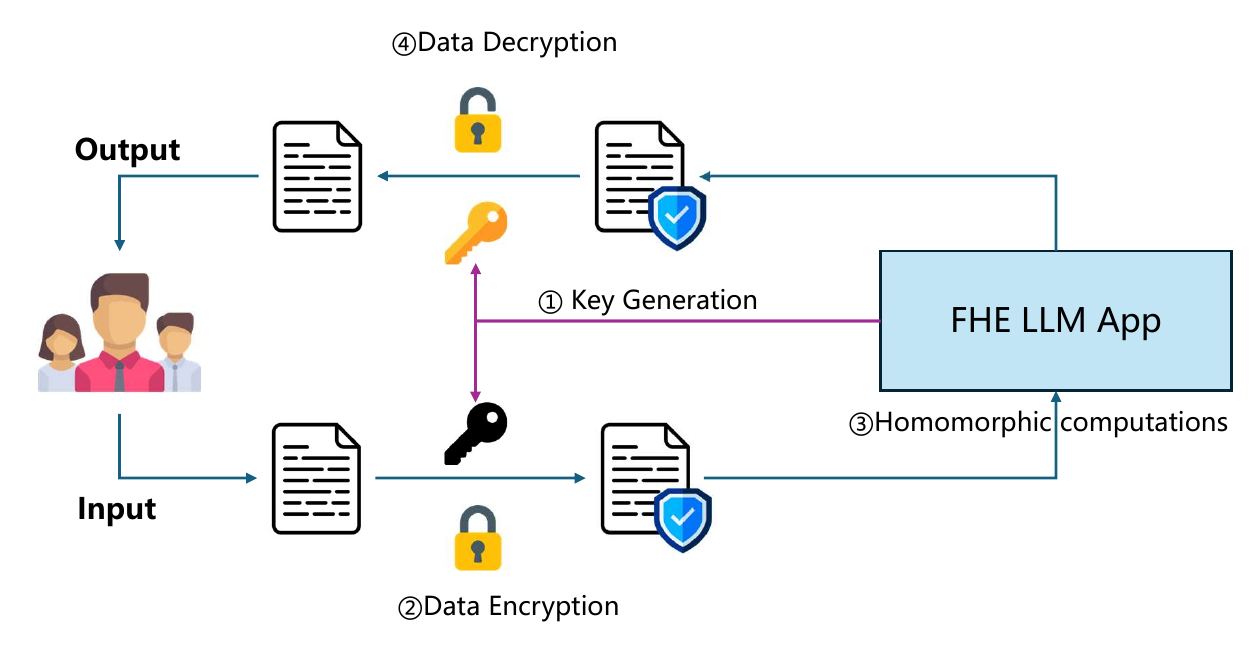}
\caption{An Example of a Local Fully Homomorphic Encrypted Large Language Model Application}
\label{F202407010018}
\end{figure}

\begin{table}[b]
\caption{Inference Performance of Self-Attention}
\label{T202407010135}
\begin{threeparttable}
\begin{tabular}{cccc}
\toprule
($d_{model},num_{head}$) & Pytorch (s) & TenSeal (s) & ConcreteML (s) \\
\midrule
(768,12) \tnote{1}               & 0.00008   & 4.50       & 182.63         \\
(3584,16) \tnote{2}            & 0.00011   & 18.83       & 663.83         \\
(3584,28) \tnote{3}           & 0.00012   & 20.16       & 722.02         \\
(4096,32) \tnote{4}         & 0.00018   & 25.60      & 866.81       \\
\bottomrule
\end{tabular}
 \begin{tablenotes}   
        \footnotesize             
        \item[1] GPT2\cite{radford2019language} and BERT\cite{devlin2018bert} use 768 vector dimensions and 12 attention heads.
        \item[2] gemma2-9b\cite{gemma_2024} uses 3584 vector dimensions and 16 attention heads.
        \item[3] Qwen2-7B\cite{bai2023qwen} uses 3584 vector dimensions and 28 attention heads.
        \item[4] LLaMA2-7B\cite{touvron2023llama} and ChatGLM3-6B\cite{team2024chatglm} use 4096 vector dimensions and 32 attention heads.
        
      \end{tablenotes}  
\end{threeparttable}
\end{table}

FHE is commonly employed in cloud scenario to safeguard users' data privacy. It allows for computations on encrypted user data, yielding decrypted results that are exactly the same as if the computations were performed on the unencrypted data\cite{acar2018survey}. We transition the FHE techniques to the LLM service for user privacy protection. \textbf{We found that Fully Homomorphic Encryption (FHE) is too heavy for LLM inference, resulting in latency increases by nearly 6 orders of magnitude compared to plaintext inference.}

As show in Figure \ref{F202407010018}, a FHE-based LLM application consists of 4 steps: 
1) it generates a pair of keys to encrypt the input and decrypt the output, respectively.
2) when the user sends content to the FHE LLM application, the encryption key is first used to encrypt the content, ensuring that the FHE-based LLM application can only receive the encrypted data.
3) the FHE-based LLM application processes the encrypted input data through homomorphic computation to derive the logically encrypted result.
4) upon the computation results being sent back to the user interface, the decryption key is utilized to decrpt the data, presenting the plain-text results on the screen.
%1),  2),  3),  4), w

We tested several Self-Attention implementations of LLM on Intel® Core™ i7-11800H processors using Pytorch\cite{paszke2017automatic} and two homomorphic computational libraries TenSeal\cite{tenseal2021} and ConcreteML \cite{ConcreteML}. The latency of the implementations is shown in Table \ref{T202407010135}. Tests were performed using inputs with a batch size of 1, a sequence length of 1, and a vector dimension of \(d_{model}\). The results show that the performance of TenSeal drops by 5 orders of magnitude compared to Pytorch, while the ConcreteML library drops by 6 to 7 orders of magnitude. This shows that the computational performance of homomorphic encryption is weak, and it is difficult to meet the demand of real-time LLM inference.

The computation speed of fully homomorphic encryption is much slower than that of ordinary computation, primarily due to its reliance on complex mathematical operations, data bloat, introduction of ciphertext noise, the need for multiple encryption and decryption processes, low algorithm efficiency, and unoptimized hardware.

\subsection{Challenges in Trusted Execution Environment}\label{sec:challenges_in_tee}
% \begin{table*}[]
% \caption{Comparison of studies on DNN inference in TEE}
% \label{T202407132057}
% \begin{tabular}{cccccc}
% \toprule
% work & Transformer Support & Methods & Edge Support & GPU Support & User Context Protection \\
% \midrule
% DarknetTZ\cite{mo2020darknetz}            & \XSolidBrush & ARM TrustZone & \Checkmark & \Checkmark & \XSolidBrush \\
% ShadowNet\cite{sun2023shadownet}          & \XSolidBrush & ARM TrustZone & \Checkmark & \Checkmark & \XSolidBrush \\
% T-Slices\cite{islam2023confidential}      & \XSolidBrush & ARM TrustZone & \Checkmark & \XSolidBrush & \XSolidBrush \\
% TransLinkGuard\cite{li2024translinkguard} & \Checkmark & Intel SGX     & \XSolidBrush & \Checkmark & \XSolidBrush \\
% \bottomrule
% \end{tabular}
% \end{table*}

Another intuitive solution to protect KV pairs is running LLM inference in the TEE, which is designed for privacy-sensitive code and data. We summarize multiple works utilizing TEE to safeguard conventional deep learning models like ResNet, VGG, and MobileNet, addressing limited memory and lack of GPU acceleration. We will discuss the inspirations from existing works and the new challenges posed by LLMs.
% Another intuitive solution to protect the KV pairs is running the LLM inference in the TEE, specially designed to execute privacy-sensitive code and data. As summarized in Table \ref{T202407132057}, there are multiple works that utilize TEE to safeguard the execution of conventional deep learning models, such as ResNet, VGG, and MobileNet, addressing the limited memory and lack of GPU acceleration support. In the following, we discuss the inspirations brought by existing works and the new challenges provided by LLM.

% TransLinkGuard\cite{li2024translinkguard} and ShadowNet\cite{sun2023shadownet} belong to one kind, do some permutation in the secure world, then use the outside work heterogeneous cores to do processing.
\begin{table}[b]
\caption{Memory layout of the Mobiles SoC's TEE}
\label{T202407072041}
\begin{threeparttable}
\begin{tabular}{ccc}
\toprule
Chips      & TZDRAM\tnote{1} (MiB) & Total DRAM (GiB) \\
\midrule
RK3399     & 32           & 4         \\
MT8173     & 30           & 2         \\
Hikey960   & 16           & 3        \\
Raaspberry Pi 3   & 15           & 1         \\
\bottomrule
\end{tabular}
 \begin{tablenotes}   
        \footnotesize             
        \item[1] TZDRAM: TrustZone DRAM.
\end{tablenotes} 
\end{threeparttable}
\vspace{-8pt} 
\end{table}

% \textbf{Incrementally feed the model layer into the TEE for model weights protection\cite{mo2020darknetz} \cite{islam2023confidential}. } DarkneTZ \cite{mo2020darknetz} and T-Slice \cite{islam2023confidential} are more concerned with model weight leakage, as effective membership inference attacks (MIAs) can disclose information about their training data. As shown in Table\ref{T202407072041} and Table\ref{T202407020937}, TZDRAM is too small for CNNs, considering that the TEE trustworthy memory is too small, DarkneTZ slices the CNN layer-by-layer enable model inference in the TEE.  T-Slice \cite{islam2023confidential}, on the other hand, run the entire model in TEE. It dynamically splits the deep learning model into units (slices) that can be executed in TrustZone's limited trusted memory without modifying the protected deep learning model. 
\textbf{Incrementally feed the model layer into the TEE for model weights protection\cite{mo2020darknetz} \cite{islam2023confidential}.} DarkneTZ \cite{mo2020darknetz} and T-Slice \cite{islam2023confidential} primarily focus on preventing model weight leakage, as effective membership inference attacks (MIAs) can reveal information about their training data. As shown in Table \ref{T202407072041} and Table \ref{T202407020937}, TZDRAM is too small for CNNs due to the limited size of TEE trustworthy memory. DarkneTZ addresses this by slicing the CNN layer-by-layer to enable model inference within the TEE. Conversely, T-Slice \cite{islam2023confidential} runs the entire model in the TEE. It dynamically splits the deep learning model into units (slices) that can be executed in TrustZone's limited trusted memory without modifying the protected deep learning model.

\textbf{Offloading computation-intensive operators to GPU devices \cite{sun2023shadownet} \cite{li2024translinkguard}.} TransLinkGuard \cite{li2024translinkguard} protects models during local inference by generating a locked model through rearranging the weights of the Transformer's fully-connected layers. During inference, TransLinkGuard rearranges the intermediate variables in the TEE to prevent model weight leakage. ShadowNet \cite{sun2023shadownet} observes that linear layers (including convolutional and fully connected layers) account for over 99\% of the weights and computation time. It outsources these linear layers to untrusted environments (including GPUs) for acceleration without leaking model weights.

% \textbf{Offloading the computation-intensive operators to heterogeneous devices \cite{sun2023shadownet} \cite{li2024translinkguard}.} TransLinkGuard \cite{li2024translinkguard} generates a locked model by rearranging the weights of the Transformer model's fully-connected layers, enabling model protection during local model inference. During the inference process, TransLinkGuard rearrange the intermediate variables in the TEE to prevent the model weights leakage. ShadowNet \cite{sun2023shadownet} finds that linear layers (including convolutional and fully connected layers) take up more than 99\% of the weights and computation time. It outsources the linear layers to untrustworthy environments (including GPUs) for acceleration without leaking the model weights. 

%It first performs inference by transforming the weights of the linear layers prior to outsourcing, and recovers the correct results by linear transformation inside the TEE, while realizing the computation of the nonlinear layers (activation layers) inside the TEE, thus ensuring data security.
%ShadowNet is a non-linear layer (activation layer), which ensures the security of the data.

\begin{table}[]
\caption{Params, memory and FLOPs of common models}
\label{T202407020937}
\begin{threeparttable}
\begin{tabular}{cccc}
\toprule
Model       & Params(M) & Mem(MiB)\tnote{3}& GFLOPS \\
\midrule
Qwen2-7B\cite{bai2023qwen}\tnote{1}   & 7070       & 30115.7          & 1680           \\
ChatGLM3-6B\cite{team2024chatglm}\tnote{1}  & 6240       & 23,794.2        & 1580           \\
LLama2-7B\cite{touvron2023llama}\tnote{1}& 6610       & 25874.1          & 1670           \\
ResNet50\cite{he2016deep}\tnote{2}   & 25.6     & 89.8             & 8.2         \\
MobileNetV2\cite{sandler2018mobilenetv2}\tnote{2} & 3.5     & 74.9             & 0.6          \\
Vgg11\cite{simonyan2014very}\tnote{2}  & 132.9    & 54.9             & 15.2    \\
\bottomrule
\end{tabular}
 \begin{tablenotes}   
        \footnotesize             
        \item[1] indicates that it is based on the Transformer large language model.
        \item[2] indicates that it represents a convolutional neural network.
        \item[3] indicates the peak memory usage during model inference.
      \end{tablenotes} 
\end{threeparttable}
\vspace{-8pt}
\end{table}
\textbf{Compared to the model weights, the KV pairs are more in need of protection}. For model privacy security of LLMs, KV pairs leakage leads to the recreation of user conversation. This is more direct and dangerous than the security risk of obtaining training data through model weights. As shown in Table \ref{table:size_of_kv_cache} illustrates that a typical user-LLM conversation spans approximately 1000 tokens. This results in the LLM generating hundreds of MiBs or even several GiBs of KV Cache. The size of this KV Cache is over 10 times larger than the model weights of a CNN, thereby making the slicing in TEE significantly more challenging.

\textbf{The lack of support for GPU acceleration in the TEE makes it challenging to efficiently perform LLM inference.} Despite having adequate memory resources in the TEE, the limitation to using only the CPU for LLM inference can lead to inefficiencies. Achieving the efficiency provided by insecure GPUs while ensuring the security of KV pairs poses a significant challenge.
%Only the CPU is available in the TEE. Running in the 

%Compared to the past TEE-protected model inference process realized by model layer slicing on CNN, the model weights of LLM shown in Table \ref{T202407020937} are three orders of magnitude higher than those of CNN, which makes such slicing too large and complex. In addition, if the optimization of AI gas pedal is missing, the inference efficiency of end-side LLM decreases significantly. Therefore, when realizing the protection of LLM user contexts on the end-side, we should follow the following design principles:

\section{Design of KV-shield}
% \deyu{Given the experiments we have done before, what kind of design we need to realize secure and efficient on-device LLM inference? 1. use a lightweight encryption scheme instead of FHE. 2. use TEE in a pipeline manner to adapt to the limited memory resources. 3. intelligently offload part of the computation to heterogeneous cores such as GPU and NPU.}

In this part, we design KV-shield to protect the original KV pairs stolen by a malicious process. We use the TEE and a simply yet effective and efficient permutation operation. We design KV-shield according to the following three principles:
\begin{enumerate}[1)]
    \item Deploy the model without modification.
    \item Keep the original KV invisible to the insecure GPUs in the REE(Rich Execution Environment).
    \item Using GPUs in REE as much as possible to improve the inference efficiency of LLM.
\end{enumerate}

\begin{figure*}[t]
\centering
\includegraphics[width=0.9\textwidth]{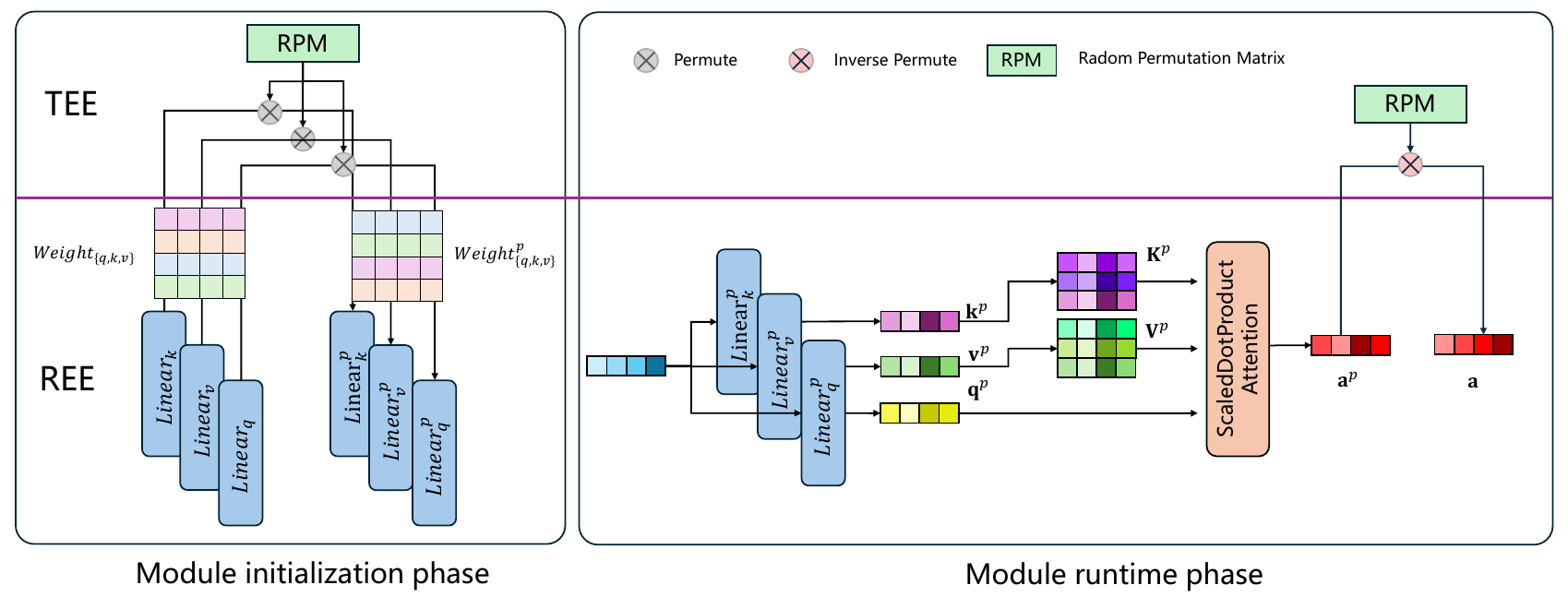}
\caption{Workflow for End-Side Protection of KV Cache}
\label{F202407151055}
\end{figure*}
\begin{figure}
\centering
\includegraphics[width=0.45\textwidth]{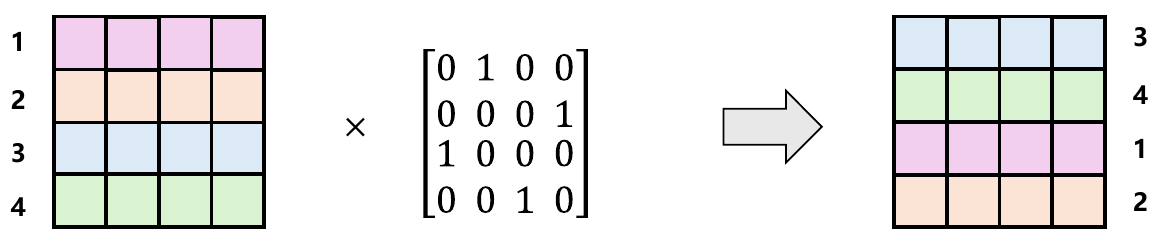}
\caption{Workflow for Matrix Permutaion}
\label{fig:permuation_workflow}
\end{figure}

\begin{table}[]
\caption{Size of KV pairs generated by LLM in single decode}
\label{table:size_of_kv_cache}
 \begin{threeparttable}
\begin{tabular}{cccc}
\toprule
Model       & $\mathrm{shape}_{KV}$ & $\mathrm{layer}_{num}$  & $\mathrm{size}_{KV}$\tnote{1}      \\
\midrule
LLaMA2-7B   & $2\times 32\times 128$ & 32        & $seq_{len}\tnote{2} \times1$ MiB   \\
ChatGLM3-6B & $2\times2\times 128$  & 28        & $seq_{len} \times 56$ KiB  \\
Qwen2-7B    & $2\times4\times 128$  & 28        & $seq_{len} \times112$ KiB \\
\bottomrule
\end{tabular}
 \begin{tablenotes}   
        \footnotesize             
        \item[1] $\mathrm{size}_{KV} =seq_{len} \times \mathrm{shape}_{KV} \times \mathrm{layer}_{num} \times 4\mathrm{B} $. The size of float32 type is 4B. 
        \item[2] The size of $seq_{len}$ is equal to the sum of the number of tokens entered by the user and the number of tokens that LLM has generated
\end{tablenotes} 
\end{threeparttable}
\vspace{-16pt}
\end{table}
\subsection{Overview}
As depicted in Figure \ref{F202407151055}, to protect the original KV pairs, we permute the weights of the linear layers in the self-attention operator. It rearranges the rows or columns of the matrices, as shown in Figure\ref{fig:permuation_workflow}.By multiplying a matrix by a 01 matrix, we can realize that the rows of the matrix are disrupted. In such a way, after the GPU performs the linear layer computations, the insecure cache stores the permuted KV pairs. The corresponding permutation matrix is stored in the TEE to ensure that attackers cannot obtain the permutation information. Finally, the results of self-attention are inversely permuted through the TEE to obtain the correct results.
\begin{enumerate}[1)]
    \item At the initialization of the LLM process, we randomly permute the weights of \(Linear_q\), \(Linear_k\), and \(Linear_v\) for each layer's self-attention module, resulting in \(Linear^{p}_q\), \(Linear^{p}_k\), and \(Linear^{p}_v\). The RPM (Random Permutation Matrix) is stored inside the TEE.
    \item When executing the self-attention module at \(Layer_i\), the input \(x\) undergoes three linear transformations (\(Linear^{p}_q\), \(Linear^{p}_k\), \(Linear^{p}_v\)) to produce the variables \(\mathbf{q}^{p}\), \(\mathbf{k}^{p}\), and \(\mathbf{v}^{p}\). The variables \(\mathbf{k}^{p}\) and \(\mathbf{v}^{p}\) are added to the KV cache, forming \(\mathbf{K}^{p}\) and \(\mathbf{V}^{p}\).
    \item The variables \(\mathbf{q}^{p}\), \(\mathbf{K}^{p}\), and \(\mathbf{V}^{p}\) are then used in the Scaled Dot-Product Attention calculation to obtain the attention results, which are sent to the TEE to recover the correct attention output.
\end{enumerate}
Our design ensures that the KV matrix stored in the insecure cache is always in the form of a permuted matrix. Even if the KV pairs are leaked, without the RPM stored in the TEE, an attacker cannot effectively recover the contextual content from the transposed KV pairs.

\subsection{Correctness and Security Analysis} \label{sec:Correctness_Analysis}
This section theoretically analyzes the correctness of the computational process depicted in Figure \ref{F202407151055}. We use \(\mathbf{RPM}\) to denote the random permutation matrix. Let \(\mathbf{x} \in \mathbb{R}^{1\times d}\), \(\mathbf{Weight} \in \mathbb{R}^{d\times d}\), \(\mathbf{RPM} \in \mathbb{R}^{d\times d}\), and \(\mathbf{RPM} \times \mathbf{RPM}^{\top} = \mathbf{I}\), where \(\mathbf{I}\) is the identity matrix.

\begin{align}
\mathbf{Weight}^{p}_{\{q, k, v\}}  &= \mathbf{Weight}_{\{q, k, v\}} \mathbf{RPM} \label{eq:shield_step1} \\
\mathbf{\{q, k, v\}}^{p}  &= \mathbf{x} \mathbf{Weight}^{p}_{\{q, k, v\}} = \mathbf{\{q, k, v\}} \mathbf{RPM}  \label{eq:shield_step4} \\
\mathbf{K}^{p} &= \begin{bmatrix}
   \mathbf{K}^{p} \\
   \mathbf{k}^{p}
   \end{bmatrix}, \quad
\mathbf{V}^{p} = \begin{bmatrix}
   \mathbf{V}^{p} \\
   \mathbf{v}^{p}
   \end{bmatrix} \label{eq:shield_step8} \\
{\mathbf{a}}^{p}  &= \mathrm{Softmax}\left(\frac{{\mathbf{q}^{p}} {{\mathbf{K}^{p}}^\top}}{\sqrt{d_k}}\right) \mathbf{V}^{p} \label{eq:shield_step9} \\
\mathbf{a} &= \mathbf{a}^{p} \mathbf{RPM}^{\top} \label{eq:shield_step10}
\end{align}

In Equation \ref{eq:shield_step1}, we permute the original weight matrix \(\mathbf{Weight}\) to \(\mathbf{Weight}^p\) using the random permutation matrix. Through these permuted weight matrices, collectively called \(\mathbf{Weight}^p\), we compute the permuted vectors \(\mathbf{q}^p\), \(\mathbf{k}^p\), and \(\mathbf{v}^p\), as shown in Equation \ref{eq:shield_step4}. 

According to our derivation, \(\mathbf{a}^p \in \mathbb{R}^{1 \times d}\). In this manner, the output attention vector \(\mathbf{a}^p\) is the permuted version of the correct attention vector \(\mathbf{a}\). Finally, within the TEE, we anti-permute \(\mathbf{a}^p\) back to \(\mathbf{a}\).

In summary, by operating on the permuted weights and inverse permuting the output attention vector, we ensure that the output of each layer remains unchanged.

\textbf{Security Analysis.} 
In KV-shield, we ensure that the plaintext of $\mathbf{K}$ and $\mathbf{V}$ are not saved in the memory of REE, thus the insecure GPUs in REE cannot process the original KV pairs directly. Even the permuted KV pairs are leaked, the attacker cannot recreate the user conversation.

%and that the inverse permutation of the Attention result is done in the TEE to ensure that the permutation information does not appear in the non-secure memory. The process of inverse substitution of Attention results is done within the TEE, ensuring that substitution information does not appear in unsafe memory.

\section{Discussion}
% \deyu{@huan, accomplish this part. Discuss the memory issue.}

\paragraph{\textbf{Feasibility of KV-shield in terms of KV protection}} Most LLM models have a model dimension $d_{model}$ of around 4096, and the current TEE memory of approximately 32MB is sufficient to accommodate part of the vector storage and computation for LLMs. TEE is sufficient to accommodate $\mathbf{a}$ to perform the permute computation.
\begin{table}[]
\caption{Overhead  of permutation }
\label{table:overhead}
\begin{tabular}{ccc}
\toprule
$d_{model}$ & Permute Weight (s) & Permute Result (s) \\
\midrule
768                         & 15.75              & 0.9                        \\
3584                        & 71.44              & 3.7                        \\
4096                        & 84.22              & 4.3                       \\
\bottomrule
\end{tabular}
\end{table}
\paragraph{\textbf{Overhead brought by matrix permutation in TEE} }
We test the efficiency of matrix permutation in the TEE on the intel 11800H in the QEMU with 16 MB TEE memory, as shown in Table \ref{table:overhead}. For ease of implementation, we implement the matrix permutation by loops. The results show that the weights and attention vectors permutation in TEE achieves the orders of seconds. To adapt to the limited TEE memory, we calculate the values of weights permutation in a block by block manner. Note that the results is just for one layer. For an entire model with over 20 layers, the latency can reach 5 minutes, which is unacceptable for users. The latency caused by vector permutation happens in the runtime phase. Although the TEE memory is sufficient for each vector permutation, the latency is still too high for real-time token generation. These results inspire us to further optimize the efficiency of KV-Shield in the future.
%However, the only problem is that every time we start the dialog process, we need to regenerate the RPM in TEE in order to allow the Attention part of the fully-connected layer weights to complete the matrix replacement operation in TEE, as shown in Table\ref{table:overhead}， We test the efficiency of our key designs on the intel 11800H again, and since the TEE memory may not be sufficient to complete the conversion in one go, we may need to step through the conversion multiple times, which may result in a long initialization time. We need to further optimize the efficiency of permutation within TEE.

\section{Conclusion}

We demonstrate that a malicious process can steal KV pairs during LLM inference on the mobile GPU and reconstruct the entire user conversation, leading to significant security vulnerabilities. To address this issue, we explore potential solutions, such as FHE and TEE, to secure on-device LLM inference. We find that FHE is too resource-intensive for on-device inference, and TEE faces limitations in both memory and computation resources. Building on these insights, we designed the KV-Shield. By permuting the weight matrix and subsequently inversely permuting the results, KV-Shield harnesses the computational power of insecure GPU accelerators while ensuring that they cannot access the original KV pairs, thus preventing data leakage. KV-Shield operates in two phases within the TEE. During the initialization phase, it shuffles the linear weights, and in the runtime phase, it reverses the permutation of the attention vectors for each self-attention module. Given the limited size of the attention vector for each module, the TEE has sufficient resources for this operation. We analyze the theoretical accuracy of the KV shield. Moving forward, we will further optimize the performance of KV-Shield for LLM inference on the device.
% We demonstrate that the malicious process can steal the KV pairs during on-device LLM inference and reconstruct the entire user conversation, leading to significant security vulnerabilities. To address this issue, we explore potential solutions, i.e., FHE and TEE, to secure the on-device LLM inference. We find that the FHE is too resource-intensive for on-device inference, and the TEE faces limitations in both memory and computation resources. Building on these insights, we deign KV-Shield. By permuting the weight matrix and subsequently inverse permute the results, KV-Shield harnesses the computational power of the insecure AI accelerators, while ensuring that they cannot access the original KV pairs, thereby preventing data leakage. It runs in two phases in the TEE. During the initialization phase, it shuffles the linear weights, while in the runtime phase, it reverses the permutation of the attention vectors for each self-attention module. Given the limited size of the attention vector for each attention module, the TEE possesses sufficient resource for this operation. We analyze the correctness of KV-shield theoretically. Moving forward, we will further optimize the security and performance of KV-shield for on-device LLM inference.
%We make the first effort to execute the on-device LLM inference in a secure and efficient manner. Towards this goal, 
\section*{Acknowledgments}

This work was supported in part by the National Key Research and Development Program of China under Grant 2022YFF0604504; in part by the National Science Foundation of China under Grant 62172439; in part by the Major Project of Natural Science Foundation of Hunan Province under Grant 2021JC0004; in part by the National Science Fund for Excellent Young Scholars of Hunan Province under Grant 2023JJ20076; and in part by the Central South University Innovation-Driven Research Programme under Grant 2023CXQD061.

 % bibliographystyle {} 
\bibliographystyle{ACM-Reference-Format}
\bibliography{main}

\end{document}